\documentclass[12pt,a4paper]{article}
\usepackage{graphicx}
\usepackage{times}
\title{\bf The chemical composition of the Orion star forming region: stars, gas and dust}
\author{S. Sim\'on-D\'iaz$^{1,2}$, M.~F. Nieva$^{3}$, N. Przybilla$^{4}$ and G. Stasi\'nska$^5$\\
\vspace{1cm}\\
\normalsize $^1$ Instituto de Astrof\'isica de Canarias, E-38200 La Laguna, Tenerife, Spain.\\
\normalsize $^2$ Departamento de Astrof\'isica, Universidad de La Laguna, E-38205 La Laguna, Tenerife, Spain \\
\normalsize $^3$ Max-Planck-Institut f\"ur Astrophysik, Karl-Schwarzschild-Str. 1, D-85741 Garching, Germany \\
\normalsize $^4$ Dr. Karl Remeis-Sternwarte \& ECAP, Sternwartstr.7, D-96049 Bamberg, Germany\\
\normalsize $^5$ LUTH, Observatoire de Paris, CNRS, Universit\'e Paris Diderot, F-92190 Meudon, France}

\date{\mbox{}}
\begin{document}
\maketitle
\pagestyle{empty}
%
%
\def\bull{\vrule height .9ex width .8ex depth -.1ex}
\makeatletter
\def\ps@plain{\let\@mkboth\gobbletwo
\def\@oddhead{}\def\@oddfoot{\hfil\tiny\bull\quad
``The multi-wavelength view of hot, massive stars''; 39$^{\rm th}$ Li\`ege Int.\ Astroph.\ Coll., 12-16 July 2010 \quad\bull}%
\def\@evenhead{}\let\@evenfoot\@oddfoot}
\makeatother
%
%
\def\beginrefer{\section*{References}%
\begin{quotation}\mbox{}\par}
\def\refer#1\par{{\setlength{\parindent}{-\leftmargin}\indent#1\par}}
\def\endrefer{\end{quotation}}
%
%
{\noindent\small{\bf Abstract:} 
We present a summary of main results from the studies performed
in the series of papers {\it ``The chemical composition of the Orion star
forming region''}. We reinvestigate the chemical composition of B-type
stars in the Orion OB1 association by means of state-of-the-art
stellar atmosphere codes, atomic models and techniques, and
compare the resulting abundances with those obtained from the
emission line spectra of the Orion nebula (M42), and recent
determinations of the Solar chemical composition.
}
%
%
\section{Introduction}

For many years, our knowledge about the chemical composition of
early-B main sequence stars in the Solar vicinity has been
characterized by two main results: (i) the derived abundances were
highly inhomogeneous (with a dispersion of up to 0.5 dex), and (ii)
the mean values indicated lower abundances than the standard set of Solar abundances
(Grevesse \& Sauval, 1998, GS98)
Some recent results have begun to change this situation. The Solar
oxygen abundance traditionally considered as a cosmic abundance
reference (GS98) was revised by Asplund et al. (2004) to a value of $\epsilon$(O) = 8.66 dex, 
0.17 dex lower than the standard
value. Przybilla, Nieva, \& Butler (2008, PNB08) have recently
analyzed a representative sample of six unevolved early B-type
stars in nearby OB associations and the field, and found a very
narrow distribution of abundances, with mean values that are larger
compared to previous works. PNB08 indicate the
importance of properly determining the atmospheric parameters
and using robust model atoms to avoid systematic errors in the
abundance determination. The study by PNB08 shows that the
chemical inhomogeneity previously found for B-type stars in the
Solar vicinity may be spurious as a consequence of those systematic errors.

\newpage

\subsection{Orion OB1 and M\,42}

The Orion complex, containing the Orion molecular cloud and the
Orion OB1 (Ori\,OB1) association, is one of the most massive active
star-forming regions in the 1 kpc centered on the Sun. Blaauw
(1964) divided Ori\,OB1 into four subgroups of stars, namely Ia,
Ib, Ic, and Id, having different locations in the sky and different ages.
Brown et al. (1994) derived mean ages of 11.4, 1.7, 4.6, and $<$1
Myr for subgroups Ia to Id, respectively. The youngest subgroup Ori\,OB1\,Id 
is associated with the Orion nebula (M\,42), the brigthest and most studied
H\,{\sc ii} region and the closest ionized nebula to the Sun in which a high
accuracy abundance analysis can be performed.
The correlation between the ages of the stellar subgroups, their
location, and the large scale structures in the interstellar medium
around Ori\,OB1 have been interpreted as features of sequential
star formation and Type-II supernovae (Reynolds \& Ogden, 1979;
Cowie et al., 1979; Brown et al., 1994).

\subsection{B-type star abundances in Orion}

Cunha \& Lambert (1992, 1994, CL92,94) obtained C, N, O, Si, and
Fe abundances of 18 B-type main sequence stars from the four
subgroups comprising the Ori\,OB1 association. They found a range
in oxygen abundances of 0.4 dex, with the highest values
corresponding to the stars in the youngest (Id and some Ic)
subgroups. In this case, the inhomogeneity in stellar abundances
(mainly oxygen and silicon) seemed to be real and coherent with a
scenario of induced star formation in which the new generation of
stars were formed from interstellar material contaminated by Type-II
supernovae ejecta.
The study by CL92,94 was based on a photometric estimation of T$_{\rm eff}$
and the fitting of the H$\gamma$ line computed from Kurucz (1979) LTE
model atmospheres to derive log\,$g$. In view of the results by PNB08
a reinvestigation of the chemical composition of B-type stars in Ori
OB1 was warranted.\\

In the series of papers {\it ``The chemical composition of the Orion starsforming region''}
we aimed at (i) investigating whether the inhomogeneity of B-type star abundances 
previously found is real, and (ii) comparing the derived abundances with recent 
determinations of the Solar chemical composition, abundances in other B-type stars in the Solar vicinity, and 
the nebular abundances derived from the emission line spectrum of the Orion nebula.
We summarize here the main results from three papers in the series.

\section{Homogeneity of O and Si abundaces in B-type stars}

In Sim\'on-D\'iaz (2010, Paper I) we investigated whether
the inhomogeneity of abundances previously found in
B-type stars in the Ori OB1 association is real or a consequence of 
intrinsic errors induced by the use of photometric indices 
to establish the stellar parameters prior to the abundance analysis.
We obtained a high quality spectroscopic data\,set
comprising 13 B-type stars in the various Ori\,OB1
associations (see the poster contribution {\it ``The IACOB spectroscopic
database of Galactic OB stars''} in these proceedings), and performed a
detailed, self-consistent spectroscopic NLTE
abundance analysis by means of the modern stellar
atmosphere code FASTWIND (Puls et al. 2005).\\

\begin{figure}[t]
\includegraphics[height =16.5cm,angle=90]{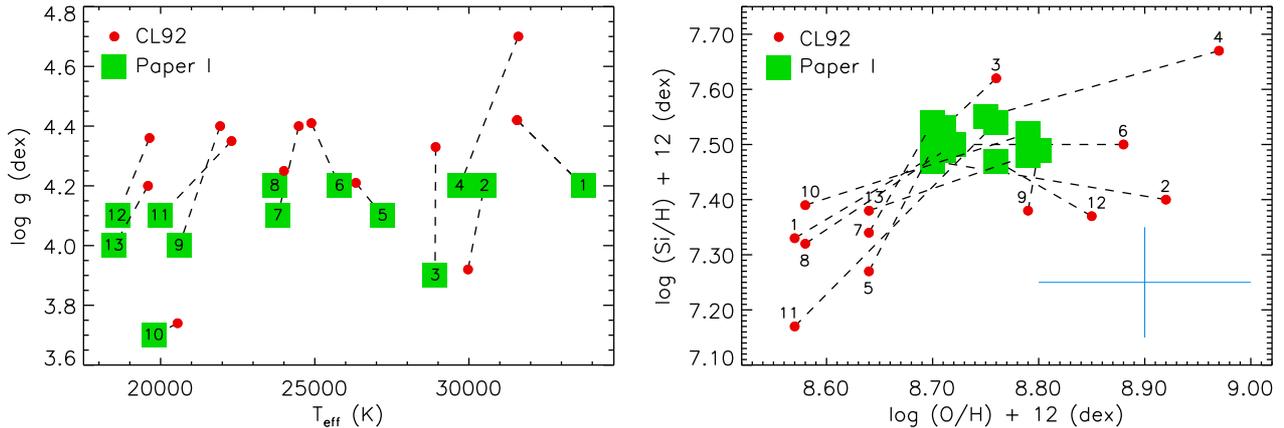}
\caption{{\small Comparison of stellar parameters, and O and Si abundances derived in Paper I
and CL92. The cyan cross represents the mean intrinsic uncertainties 
for O and Si abundances. Numbering follows the order in Table 2 from the original paper.\label{fig_1}}}
\end{figure}

Main results of this study are summarized in Figure \ref{fig_1} (from Paper I) 
and Table \ref{tab_1}. We detected systematic errors in the stellar parameters determined 
previously which affected the derived abundances. Once these errors were accounted 
for, we find a high degree of homogeneity in the O and Si abundances 
for stars in the four Ori\,OB1 subgroups. The dispersion of abundances for 
the 13 stars is smaller than the intrinsic uncertainties (0.10 and 0.08 dex for
O and Si, respectively).

The mean oxygen and silicon abundances agree with those resulting from a similar 
analysis of a representative sample of unevolved early B-type stars in the Solar 
neighbourhood (PNB08). Both results indicate that abundances derived from these stellar objects are
more homogeneous and metal rich than previously though.\\

We also compared the O and Si stellar abundance in Ori\,OB1 with those
obtained for the Sun during the epoch of the "Solar crisis". The O 
abundances in our sample of stars in Ori\,OB1 lay in the middle of 
all these values. In view of the present-day results the only thing we 
can say is that oxygen abundances in the Sun and B-type stars in the 
solar vicinity are the same within the uncertainties. However, we 
consider too premature to launch any firm conclusion or hypothesis
about the chemical evolution of the local interstellar medium during the
lifetime of the Sun. Silicon abundances are also very similar (contrarily
to what was previously found from the study of B-type stars).

\begin{table}[t]
\begin{minipage}{8cm}
\centering
\caption{{\small
Mean O and Si abundances derived from stars in each of the subgroups in Ori\,OB1. 
The global mean abundances are also indicated (FASTWIND analysis, Paper I).\label{tab_1}}}
\vspace{0.5cm}
\begin{tabular}{lcc}
\hline
\hline
Group            & $\epsilon$(O)   & $\epsilon$(Si)  \\
\hline
Ia & 8.77$\pm$0.04 & 7.50$\pm$0.02\\
Ib & 8.70$\pm$0.01 & 7.53$\pm$0.01\\
Ic & 8.74$\pm$0.04 & 7.50$\pm$0.03\\
Id & 8.72$\pm$0.04 & 7.51$\pm$0.06\\
\hline
Global & 8.74$\pm$0.04 & 7.51$\pm$0.03\\
\hline
\hline
\end{tabular}
\end{minipage}
\hfill
\begin{minipage}{8cm}
\centering
\caption{{\small
Mean C, N, Ne, Mg and Fe abundances from B stars in Ori\,OB1 (ADS analysis, Paper II).
Intrinsic uncertaintites from individual star analyses are indicated in brackets.\label{tab_2}}}
\vspace{0.5cm}
\begin{tabular}{lc}
\hline
\hline
Element            & Mean abundance  \\
\hline
C  & 8.35$\pm$0.03 (0.09)\\
N  & 7.82$\pm$0.07 (0.09)\\
Ne & 8.09$\pm$0.05 (0.09)\\
Mg & 7.57$\pm$0.06 (0.03)\\
Fe & 7.50$\pm$0.04 (0.10)\\
\hline
\hline
\end{tabular}
\end{minipage}
\end{table}

\begin{figure}[h]
\centering
\includegraphics[height =16.5cm,angle=90]{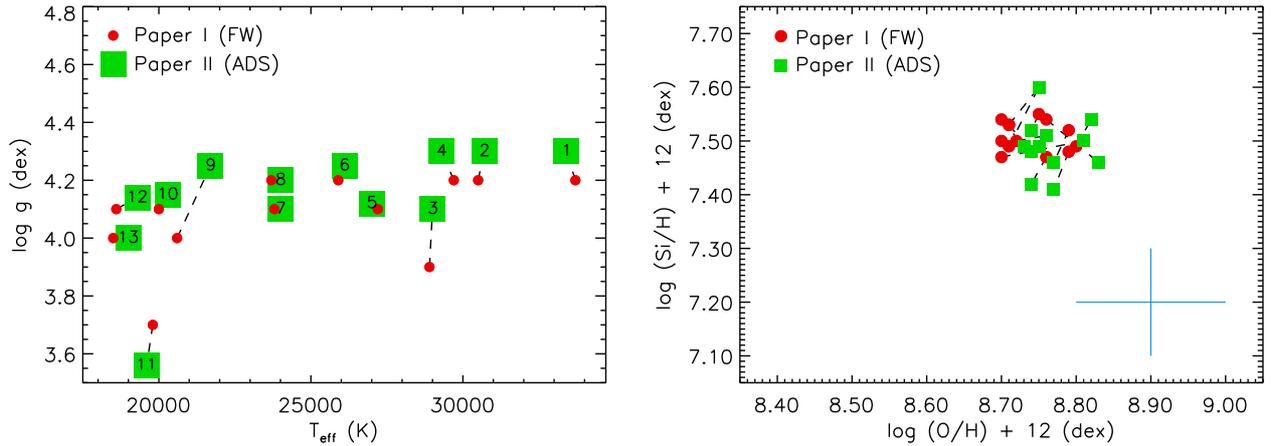}
\caption{{\small Comparison of stellar parameters, and O and Si abundances derived in Papers
I and II (FW and ADS analyses, respectively). The cyan cross represents the mean intrinsic uncertainties 
for O and Si abundances. Same numbering as in Fig. \ref{fig_1}.\label{fig_2}}}
\end{figure}

\section{C, N, Ne, Mg, and Fe abundances in B-type stars revisited}

In view of the results found in Paper I, a
reinvestigation of other stellar abundances was
warranted. In Nieva, Sim\'on-D\'iaz \& Przybilla (Paper
II, in preparation) we derive C, N, Ne, Mg and Fe
NLTE abundances for the stars analyzed in Paper I,
following the procedure described in Nieva \& Przybilla (2007, 2008).
The strategy (ADS analysis) is based on a similar self-consistent spectroscopic
analysis, but using ATLAS9 (Kurucz 1993) +
DETAIL+SURFACE (Giddings 1981; Butler \& Giddings
1985; plus recent updates).\\

We first compare the stellar parameters, O and Si abundances resulting from 
the FASTWIND (FW) and ADS analyses to be sure that both techniques provide 
similar results (see Figure \ref{fig_2}). We find very good agreement within 
the uncertainties. 
Then we obtain abundances for C, N, Ne, Mg, and Fe by means of the ADS
analysis (see final mean values from the analysis of 13 stars in the various
subgroups from Ori\,OB1 in Table \ref{tab_2}). Again, the dispersion of abundances for 
the 13 stars is smaller than the intrinsic uncertainties, indicating a high degree of
homogeneity in the present-day abundances of these elements in the Orion star forming region
(as derived from B-type stars).

\section{Stars, gas and dust: the abundance discrepancy conundrum}

\begin{figure}[h]
{\scriptsize
\begin{minipage}{8cm}
\centering
\includegraphics[height=8.cm, angle=90]{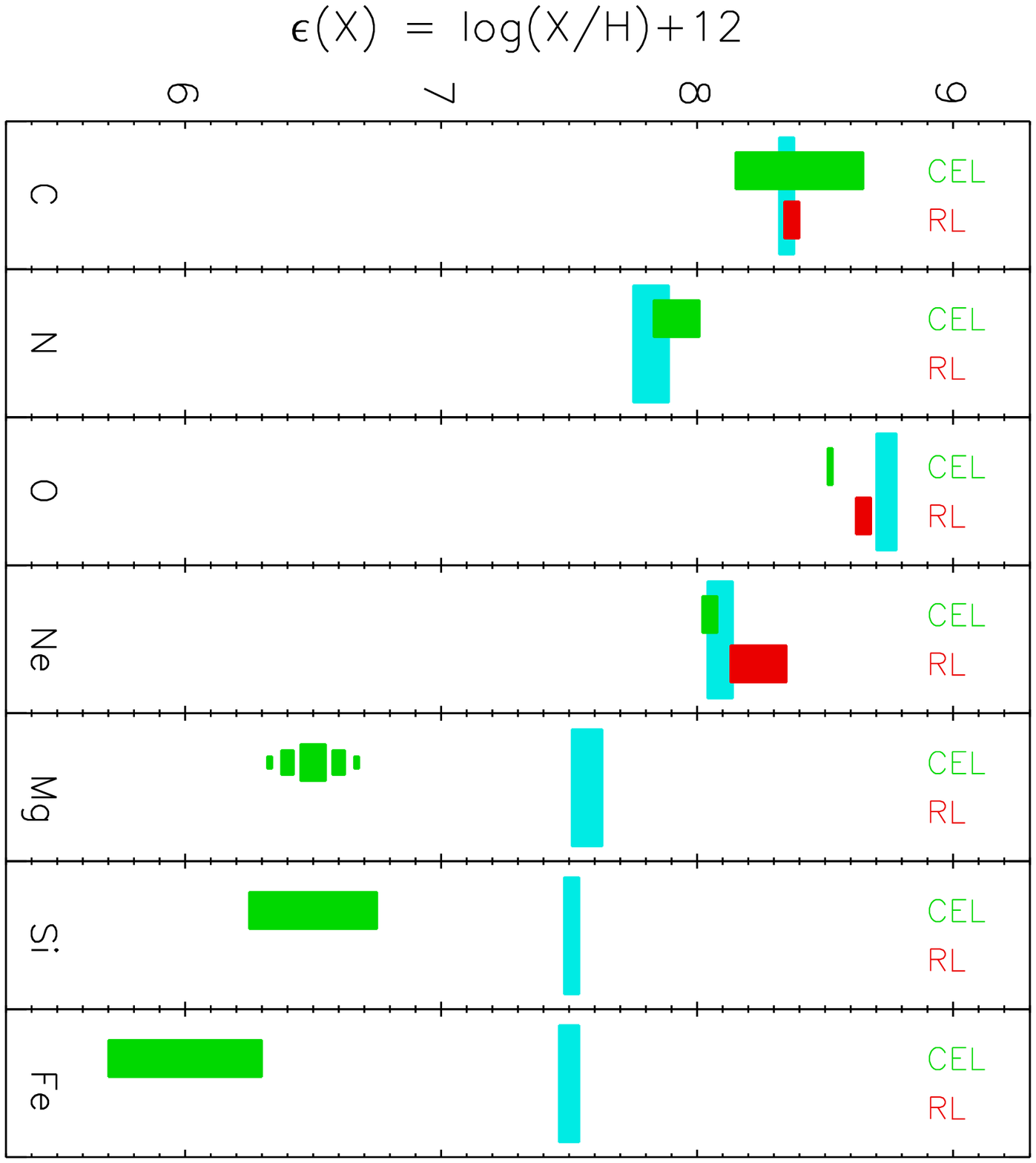}
\caption{\small{Compararison of stellar (cyan) and
nebular abundances (green and red). The heigh of
the boxes represents the uncertainties. A
more detailed comparison (also including dust
depletion) for oxygen is presented in Fig. \ref{fig_4}.} \label{fig_3}}
\end{minipage}
\hfill
\begin{minipage}{8cm}
\centering
\includegraphics[height=8.cm , angle=90]{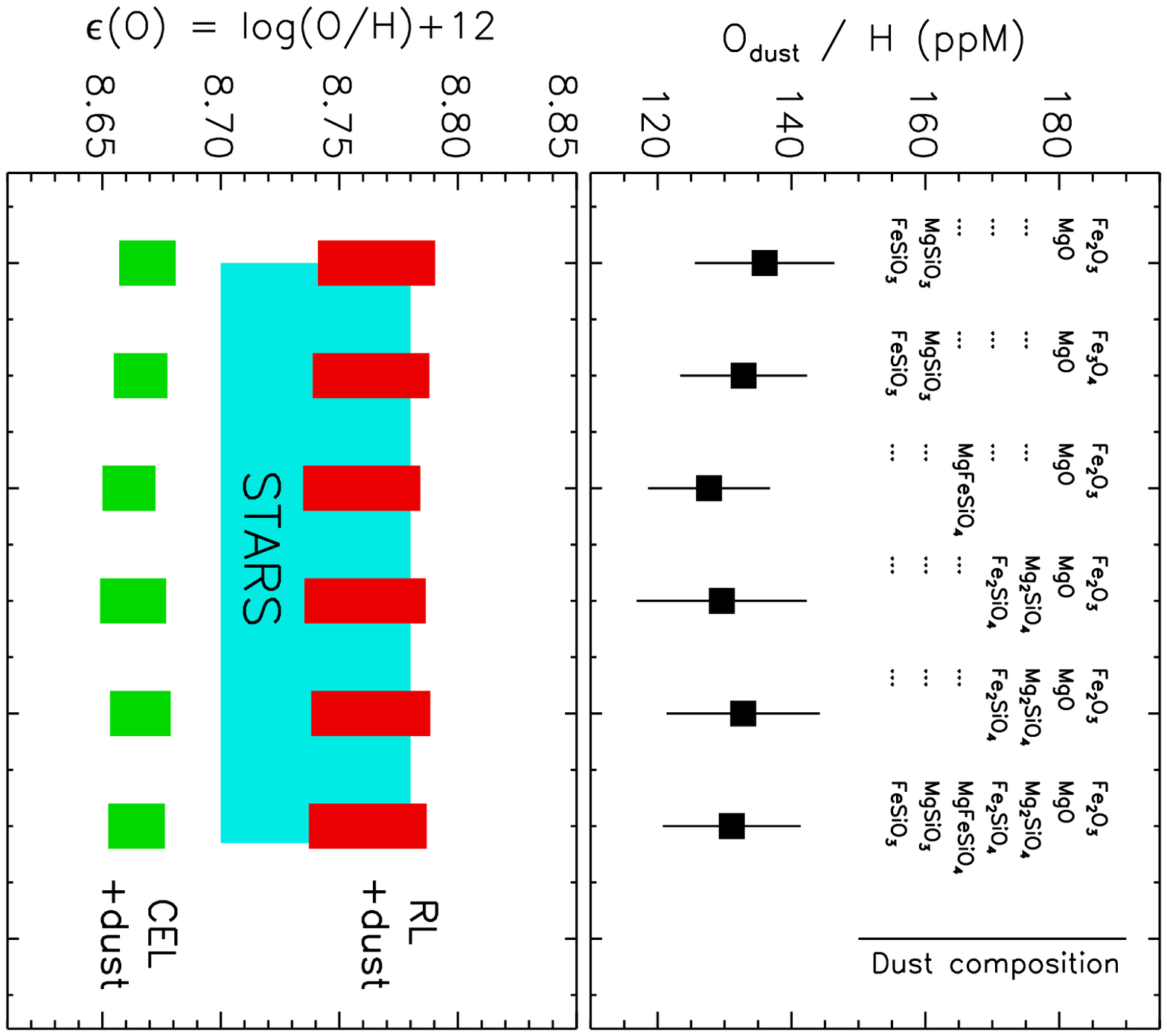}
\caption{{\small Our study leads us to the conclusion that
the oxygen abundances derived in the Orion nebula
from CELs are incompatible with the abundances
derived in the B stars (once the effects of depletion
in dust grains are taken into account) while RL
oxygen abundances are in better agreement.\label{fig_4}}}
\end{minipage}
}
\end{figure}

In Sim\'on-D\'iaz \& Stasi\'nska (Paper III, submitted), we reexamined the nebular 
abundance discrepancy problem which arises from the use of recombination lines (RL) 
on the one hand and collisionally-excited lines (CEL) on the other (see Garc\'ia-Rojas 
\& Esteban 2007, and references therein) 
in the light of the high quality abundance determinations for B-type stars performed in Papers I and II. 
We reevaluated the CEL and RL  abundances of several elements in the Orion nebula
using the spectroscopic data\,set presented by Esteban et al. (2004) and
ionization correction factors (icsf) for unseen ions obtained by means of a photonization
model suited for our purposes. The derived nebular abundances are compared with the mean
stellar abundances obtained in Papers I and II (see Fig. \ref{fig_3}). In particular, for 
the case of oxygen, we estimated the amount of oxygen trapped in dust grains for several 
scenarios for dust formation and compare the resulting gas+dust nebular abundances with the 
stellar abundances (see Fig. \ref{fig_4}).

\newpage

%
%
%
%
\footnotesize
\beginrefer
\refer  Asplund, M., Grevesse, N., Sauval, A.~J., et al.\ 2004, A\&A, 417, 751 

\refer Blaauw, A.\ 1964, ARA\&A, 2, 213 

\refer Brown, A.~G.~A., de Geus, E.~J., \& de Zeeuw, P.~T.\ 1994, A\&A, 289, 101

\refer Butler, K., \& Giddings, J. R.\ 1985, in Newsletter of Analysis of Astronomical Spectra, No. 9 (Univ. London)

\refer Cowie, L.~L., Songaila, A., \& York, D.~G.\ 1979, ApJ, 230, 469

\refer Cunha, K., \& Lambert, D.~L.\ 1992, ApJ, 399, 586 

\refer Cunha, K., \& Lambert, D.~L.\ 1994, ApJ, 426, 170 

\refer Esteban, C., Peimbert, M., Garc{\'{\i}}a-Rojas, J., et al.\ 2004, MNRAS, 355, 229

\refer Garc{\'{\i}}a-Rojas, J., \& Esteban, C.\ 2007, ApJ, 670, 457 

\refer Giddings, J. R.\ 1981, Ph.D. Thesis (Univ. London)

\refer Grevesse, N., \& Sauval, A.~J.\ 1998, Space Sci. Rev., 85, 161

\refer Kurucz, R.\ 1979, ApJS, 40, 1

\refer Kurucz, R.\ 1993, CD-ROM No. 13 (SAO, Cambridge, MA)

\refer Nieva M.F., Przybilla N.\ 2007, A\&A, 467, 295

\refer Nieva M.F., Przybilla N.\ 2008, A\&A, 481, 199

\refer Przybilla, Nieva \& Butler\ 2008, ApJ, 688, L103

\refer Puls, J., Urbaneja, M.~A., Venero, R., et al.\ 2005, A\&A, 435, 669

\refer Reynolds, R.~J., \& Ogden, P.~M.\ 1979, ApJ, 229, 942

\refer Sim\'on-D\'iaz 2010, A\&A, 510, A22

\refer Sim\'on-D\'iaz \& Stasi\'nska, A\&A, submitted

\endrefer           
\end{document}